\documentclass[11pt]{amsart}
\usepackage{geometry}                
\usepackage{graphicx}
\usepackage{amssymb}
\usepackage{epstopdf}
\usepackage{amsmath}
 \usepackage{amsmath}
    \usepackage{amssymb}
    \usepackage{amsthm}
    \usepackage{amsfonts}
    \usepackage{braket}

\usepackage{color}
\usepackage{pstricks}
\usepackage{graphicx}
\usepackage{slashed}

\title{How to Split the Electron in Half}
\author{Gordon W. Semenoff}
\address{Department of Physics and Astronomy, University of British Columbia, 6224 Agricultural Road, Vancouver, British Columbia, Canada V6T 1Z1}
\thanks{This work is supported by NSERC of Canada.}

\begin{document}

 \begin{abstract} 
 This essay is a tribute to Professor Roman Jackiw on the occasion of his eightieth year.  
 It discusses some ideas about Fermion zero modes and fractional charges and quantum entanglement. 
 \end{abstract}
\maketitle

 \tableofcontents

\section{Prologue}

 This is an essay in tribute to Professor Roman Jackiw on the occasion of his 80th year.  The beauty and the originality of Roman's scientific work inspired many   and it continues to form the bedrock of whole lines of investigation.  I am one of the inspired.  I offer this essay about some of my modest thoughts about one of the many interesting ideas of the Jackiw-Rebbi era \cite{Jackiw:1975fn}, the connections between topology, Fermion zero modes and fractional charges. This was one of the first relationships between the topology of  states of quantized fields and their physical properties \cite{Jackiw_Dirac} \cite{Niemi:1984vz}.  Of course, like all great ideas, it inspired more in its aftermath.  In this case, it is part of the nexus of ideas which has developed  into the subject of topological insulators, one of the most active and interesting areas of modern theoretical physics.    
 
 What I will discuss is a particular example of a topological insulator, an exceedingly simple one in fact, 
 where the peculiarities of the system result in quantum states with unusual properties. I will not review the topological aspects of the models, as there are already a number of beautifully explained overviews \cite{sshreview_2} \cite{sshreview_1}. I will rather concentrate on the eletronic properties of some simple models.

\section{Tight binding model}

\begin{figure}[h!]
\centerline{\includegraphics[scale=1.2]{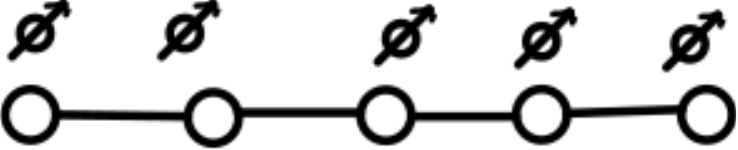}}
\caption{\small  We will consider a one-dimensional chain of atoms with an odd number of sites.  The electron density is one electron per site.  The electrons each have two spin sites.  We will assume that the coupling between the atomic and electron spins is negligible so that the system has a symmetry under the rotation of electronic spins.     }
\label{figure1}
\end{figure}

  Let us consider a hypothetical one-dimensional system consisting of an open chain of sites.  We shall assume that this chain can be occupied by electrons to a maximum density of two electrons per site.   
To a first approximation, the quantum states of the electrons are  single one-particle bound states localized at each of the sites of the chain. We will assume that the sites are charged so that the system with a density of one electron per site is charge neutral.  The electrons have spin 1/2 and the maximum density of the chain of two electrons per site is achieved by the localized electrons taking up each of the two spin states.  The charge neutral state of a chain is depicted in figure \ref{figure1}.  

The tight-binding model begins with the assumption that the electrons reside in orbitals which are localized at the lattice sites as we have described.  The
dynamics is simply taking into account the fact that an electron in one orbital has some probability of tunnelling to a neighbouring orbital.  If we put a single electron at a site on the chain, its wave-function would diffuse until, typically, it occupied the entire system.  

The tight-binding Hamiltonian for the many-electron system is gotten by first introducing creation and annihilation operators for the tight-binding states. 
The electron  bound to the $n$'th site and with spin polarization $\sigma$ is created  by $a^\dagger_{\sigma, n}$ and annihilated by $a_{\sigma, n}$.
These operators have the anti-commutation relations
\begin{align}
&\left\{ a_{\sigma_1,n_1}, a_{\sigma_2,n_2}^\dagger\right\}=\delta_{n_1n_2}\delta_{\sigma_1\sigma_2}
\\
&\left\{ a_{\sigma_1,n_1}, a_{\sigma_2,n_2} \right\}=0~,~
\left\{ a^\dagger_{\sigma_1,n_1}, a_{\sigma_2,n_2}^\dagger\right\}=0
\end{align}
The Hamiltonian is
\begin{align}\label{tight_binding_Hamiltonian}
H=\sum_{n=1}^{N-1}\sum_{\sigma=\uparrow,\downarrow}
\left( t_n a_{\sigma ,n+1}^\dagger a_{\sigma, n}+t_n^* a_{\sigma ,n}^\dagger a_{\sigma, n+1}\right)
\end{align}
where $t_n$ is the probability amplitude for an electron to tunnel from position $n$ to position $n+1$.   The amplitude for tunnelling from $n+1$ to $n$ is $t^*_n$.  For now, we have assumed that these amplitudes are independent of the electron spin states (we will change this assumption later). 
Also, for the moment, apart from the assumption that all of the hopping amplitudes are non-zero, we will make no further assumptions.   We note that, by redefining the phases of the creation and annihilation operators, we can make all of the parameters $t_n$ real and positive. We shall hereafter assume that we have done that.

We shall not be overly specific about how such a chain could be realized in nature.  An important example is a conducting polymer such as polyacetylene. 
which consists of a chain of Carbon atoms bound together by strong covalent bonds.   In that case, we would  assume that the electrons which are buried deep in atomic shells  are so tightly bound that they do not contribute to the electronic properties of the material.  A Carbon atom has four valence electrons
which are less tightly bound.  Two of the valence electrons form the strong covalent bonds with the two Carbon neighbours of each atom.  A third electron forms a  bond with a hydrogen atom which is bound to each Carbon atom.  The fourth electron on each atom is more loosely bound and it plays the role of our dynamical electron.  We will assume that this single valence electron is entirely responsible for the electronic properties of the system.  In particular, a model of trans-polyacetylene is described in this tight-binding  approximation is called the Su-Schrieffer-Heeger  (SSH) model  
\cite{Su:1979ua}-\cite{Jackiw:1981wc}.  In that model, the strength of the hopping parameter 
varies with position, $t_{2n}=t$, $t_{2n+1}=t'$ and $|t|\neq |t'|$. (We review a complete solution of the SSH model in Appendix A.)
 Another option for engineering such a chain is to use cold atoms in  an optical lattice.  There are other ideas, such as linear arrays of Josephson junctions or quantum dots that can be modelled by the SSH Hamiltonian.  

In the Heisenberg picture, the tight binding model equation of motion for the creation and annihilation operators is found by
taking the commutator
\begin{align}
i\hbar \frac{d}{dt}a_{\sigma, n} = \left[ a_{\sigma,n},H\right]
\end{align}
which yields the differential-difference equation
\begin{align}
i\hbar \frac{d}{dt}a_{\sigma ,n}= 
t_{n-1} a_{\sigma ,n-1}+t_n a_{\sigma, n+1}
\end{align}
The eigenvalue equation for the levels of the single-particle Hamiltonian can be written as a
matrix equation
\begin{align} \label{schroedinger_equation}
E\phi _{E, n}= 
t_{n-1} \phi_{E, n-1}+t_n \phi_{E, n+1}
\end{align}
which we could write as $h\phi_E=E\phi_E$
where $\phi_E$ is a vector with an amplitudes $\phi_{E,n}$, $n=1,2,...,N$, or more explicitly
\begin{align}
\phi_E = 
\left[ \begin{matrix}\phi_{E,1}\cr\phi_{E,2}\cr .\cr.\cr \phi_{E,N}\cr\end{matrix}\right]
\label{matrix_phi}
\end{align}
and 
\begin{align}
h=\left[ \begin{matrix}0&t_1&0&0& 0 & \ldots &\ldots\cr
t_1&0&t_2&0& 0 & \ldots & \ldots \cr
0 & t_2 & 0 & t_3 & 0 & \ldots & \ldots \cr
0 & 0 & t_3 & 0 & t_4 & \ldots & \ldots \cr \ldots  &  \ldots 
 &  \ldots  & \ldots & \ldots & \ldots &  \ldots \cr
 \ldots  &  \ldots 
 &  \ldots  & \ldots & \ldots & \ldots &  \ldots \cr
 \ldots  &  \ldots 
 &  \ldots  & \ldots & \ldots & 0 & t_{N-1}  \cr
 \ldots  &  \ldots 
 &  \ldots  & \ldots & \ldots &t_{N-1} &  0 \cr
 \end{matrix}\right]
\label{matrix_h}\end{align}
is the single-particle Hamiltonian.  The eigenvalues of this Hermitian matrix are the energy levels of the system.  
For the most part, we shall not need the details of the  spectrum of the Hamiltonian beyond a few observations.   
The question most important to us, about whether there are zero modes will be discussed in the next section.

\section{Fermion zero modes}

First of all, we note the rather remarkable fact that, without any further specification of the hopping amplitudes, $t_n$, the 
spectrum of the Hamiltonian has particle-hole symmetry.  To see this, we note that there is a matrix
\begin{align}
\left[ \begin{matrix}1&0&0&0& 0 & \ldots &\ldots\cr
0& -1 & 0 &0& 0 & \ldots & \ldots \cr
0 & 0 & 1 & 0 & 0 & \ldots & \ldots \cr
0 & 0 & 0 &-1 & 0 & \ldots & \ldots \cr \ldots  &  \ldots 
 &  \ldots  & \ldots & \ldots & \ldots &  \ldots \cr
 \ldots  &  \ldots 
 &  \ldots  & \ldots & \ldots & \ldots &  \ldots \cr
 \ldots  &  \ldots 
 &  \ldots  & \ldots & \ldots & \ldots &0 \cr
 \ldots  &  \ldots 
 &  \ldots  & \ldots & \ldots &0 & \pm 1 \cr
 \end{matrix}\right]
\end{align}
where, whether the last entry is $+1$ or $-1$ depends on whether $N$ is odd or even, respectively.  
This matrix has the property that $\Gamma=\Gamma^\dagger$, $\Gamma^2=I$ and
\begin{align}
\Gamma h \Gamma =-h
\end{align}
Its existence implies that, if we managed to find an eigenvector of the Hamiltonian with eigenvalue $E$,
\begin{align}
h\Phi_E=E\phi_E
\end{align}
then we can find another eigenvector of the Hamiltonian with eigenvalue of opposite sign, $-E$, as
\begin{align}
h(\Gamma \Phi_E)=-E(\Gamma\phi_E)
\end{align}
This particle-hole conjugation, when operating on a wave-function $\phi_{E,n}$ which solves equation (\ref{schroedinger_equation}) is
the transformation
\begin{align}\label{particle_hole_conjugation}
\phi_{-E,n}~=~(\Gamma\phi)_{E,n}~=~(-1)^n\phi_{E,n}
\end{align}
 Which changes the sign of the wave-function on the odd sub-lattice.  This leads to a discrete symmetry of the many-Fermion system that is described
 by the Hamiltonian $H$ in equation (\ref{tight_binding_Hamiltonian}). 
 The transformation of the the creation and annihilation operators   is
 \begin{align}
 a_{\sigma ,n}\to (-1)^n a_{\sigma,n}^\dagger
 ~,~~a_{\sigma ,n}^\dagger \to (-1)^n a_{\sigma,n}
 \end{align}
 The Hamiltonian of the many-Fermion system, $H$,  is invariant under this transformation. 

The spectrum of the single-particle Hamiltonian $h$ thus has particle-hole symmetry, in that for  each eigenvector $\phi_E$ with
positive eigenvalue $E$ there is an eigenvector $\Gamma\phi_E$ with negative eigenvalue $-E$. 
Moreover, the number of eigenvalues (per spin state)  is equal to the dimension of the quantum Hilbert space, which is $N$, the number of sites in the chain. 

If $N$ is even, 
the pairing of positive and negative energy modes  implies a one-to-one matching of the positive and negative eigenvalues.  If they exist,  zero eigenvalues must also be paired, and they must  be even in number. (We will show shortly that they do not exist at all for the even-sited chain.)

\begin{figure}[h!]
\centerline{\includegraphics[scale=0.8]{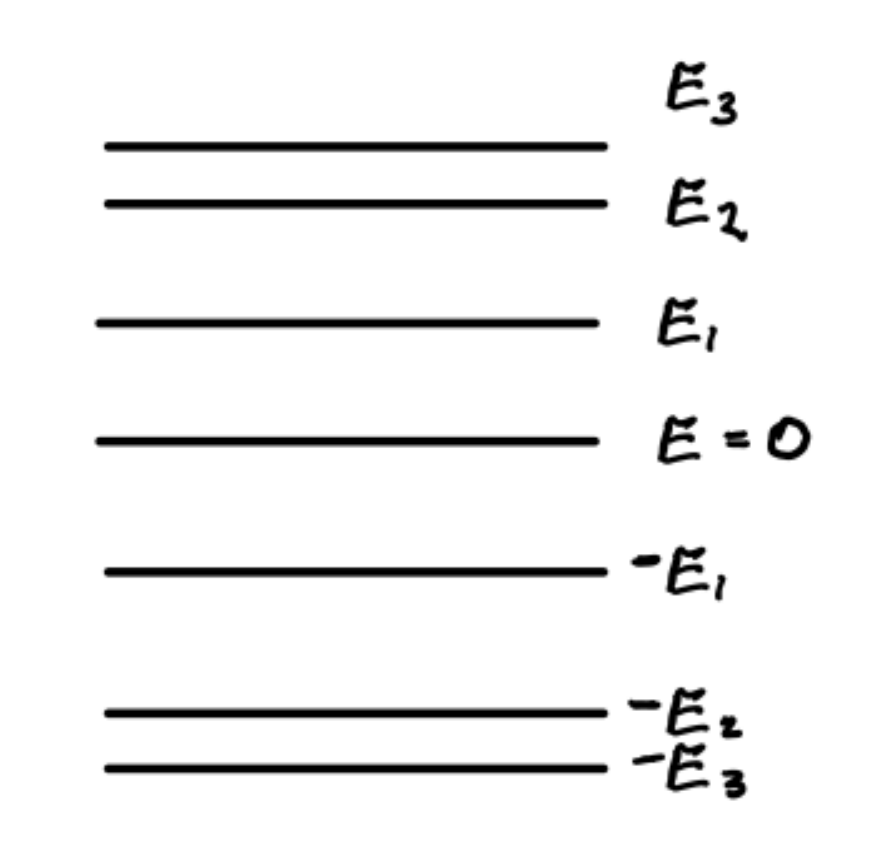}}
\caption{\small   {\it The world's simplest index theorem.}  When the number of energy levels is odd and when the spectrum is particle-hole symmetric so that there is a one-to-one mapping of positive energy levels onto negative energy levels, there must be an odd number of zero energy levels.  Generically, one would expect that perturbations of the Hamiltonian which preserve the particle-hole symmetry could lift the degeneracy of the zero modes, but this could only happen in positive-negative energy pairs, so an odd number of zero modes must remain.  We will show by explicitly examining the zero mode equations for the tight-binding problem that the odd sited chain alway has exactly one zero mode and the even sited chain can have no zero modes at all.}
\label{figure1.5}
\end{figure}

However, 
if $N$ is odd, this one-to-one matching is not possible and the particle-hole 
symmetry of the spectrum has an interesting consequence.  The positive and negative energy levels
are paired and there must be an odd one out. That odd one out must be unpaired and it must reside precisely at the centre of the spectrum, at $E=0$.  This state is what we shall call the  ``zero mode''. This circumstance is depicted in figure \ref{figure1.5}.  This argument actually establishes that there must be an odd number of zero modes.  In the following we will show that there can be only one zero mode for the odd-sited chain.  (Of course the electron state that results from this one zero mode will still have the two-fold spin degeneracy.)

At this point, we do not know where the zero mode wave-function of the odd-sited chain has support.  We also do not know to what extent the zero mode is isolated from the non-zero modes.   These details depend on the specific values of  the hopping amplitudes, $t_n$. 

\begin{figure}[h!]
\centerline{\includegraphics[scale=1]{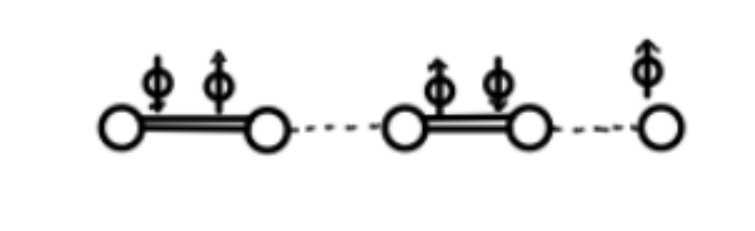}}
\caption{\small   The Su-Schrieffer-Heeger chain which is depicted here has alternating strong and weak bonds.  The electron pairs participating in the strong bonds are sufficiently bound to make a gap in their spectrum.  When there are an odd number of sites, one edge of the chain has a strong bond and the opposite edge fo the chain has a weak bond.  The odd electron out resides at the edge of the chain which has a weak bond and it occupies the zero mode in the electron energy spectrum.  When the electron has spin, this mode can either be empty, singly occupied or doubly occupied.  These states of the many-Fermion Hamiltonian are thus four-fold degenerate, although the states have differing charges and spins }
\label{figure2}
\end{figure}

There are explicit
examples, such as the SSH model \cite{Su:1979ua}-\cite{Jackiw:1981wc}, depicted in figure \ref{figure2}, where  
the non-zero
modes have a gap, which allows the zero mode, which appears at the centre of the gap, to be well isolated from the rest of the spectrum.  In that case, for the odd-sited SSH chain, there is a single zero mode with  support that is concentrated at one or the other edges of the chain.

The other extreme is the simple translation invariant   chain where all of the hopping amplitudes are equal, $t_n=t$ for all $n$.  
In that case,  the zero mode wave-function vanishes on the even sites and it is uniformly distributed on the odd sites.  However, it is not well-isolated from the rest of the spectrum since the lowest non-zero modes have energies are $\pm 2t\sin\frac{\pi}{N+1}$
which can be very small if $N$ is large.

Note that we have shown that a zero mode must exist in the odd-sited system.  
Let us examine the equation that the zero mode must obey, the 
eigenvalue equation (\ref{schroedinger_equation}) with $E=0$, 
\begin{align}\label{zero_mode_equations}
t_{2n-1} \phi_{0 ,2n-1}+t_{2n} \phi_{0 ,2n+1}=0~~,~~t_{2n} \phi_{0 ,2n}+t_{2n+1} \phi_{0, 2n+2}=0
\end{align}
These equations have the special property that they only couple the wave-functions on even sites with even sites and odd sites with odd sites. 

We can re-organize them as
\begin{align}\label{zero_mode_equations_0}&
t_1\phi_{0,2}=0~,~t_{N-1}\phi_{0, N-1}=0 \\
\label{zero_mode_equations_1}
 &\phi_{0 ,2n+2} =-\frac{t_{2n} }{t_{2n+1}}\phi_{0, 2n} n=1,2,...,(N-3)/2
\\
\label{zero_mode_equations_2}
 &\phi_{0, 2n+1}  =-\frac{t_{2n-1} }{t_{2n}}\phi_{0, 2n-1}~~,~~ n=1,2,...,(N-3)/2
\end{align}
The equations $t_1\phi_{0,2}=0$ and $t_{N-1}\phi_{0 ,N-1}=0$ in (\ref{zero_mode_equations_0}) 
are easy to see from the first and last row of the matrix equation $h\psi_0=0$
with $h$ given in equation (\ref{matrix_h}). 
The result of these equations for the odd-sited chain is that the  solution of equation (\ref{zero_mode_equations_1})  is  $\phi_{0,2n}=0$, that is, the zero mode wave-functions vanish on the 
even sites.  Then equation (\ref{zero_mode_equations_2}) has a unique solution 
\begin{align}\label{zero_mode_wavefunction}
\phi_{0 2n+1}=\left(-\frac{t_{2n-1}}{t_{2n}} \right)\left(-\frac{t_{2n-3}}{t_{2n-2}} \right)\left(-\frac{t_{2n-5}}{t_{2n-4}} \right)
\ldots  \left(-\frac{t_{3}}{t_{4}} \right)\left(-\frac{t_{1}}{t_{2}} \right)\phi_{0 1}  
\end{align}
which tells us that there is precisely one non-degenerate  zero mode and that its wave-function has support
only on the even sub-lattice. Moreover, we know its wave-function (\ref{zero_mode_wavefunction}) explicitly.


In passing, we notice  that, if the number of lattice sites were even, rather than odd, the equations $\phi_{0 ,2}=0$ and $\phi_{0,N-1}=0$
put the wave-function to zero on both the even and the odd sub-lattices, and there are no solutions for zero modes at all.  Thus, the even-sited chain can have no zero modes. This statement might not be very important for a long, almost translation invariant chain where there can typically be non-zero modes could have very small energies, exponentially small in the length of the chain.  However, for a short or meso-scopic chain, the fact that there can be no zero modes on an even-sited chain and precisely one zero mode on an odd-sited chain, no matter what the hopping parameters, could be very useful information.

In addition, what is very interesting about the odd-sited chain is that, if one could engineer the $t_n$'s appropriately, one could determine the support of the zero mode. For example, if $|t_{\rm odd}|<|t_{\rm even}|$, the zero mode has support near one end of the chain, with $\phi_1$ being the largest component.
On the other hand, if $|t_{\rm odd}/t_{\rm even}|\sim 1$ the support of the zero mode wave-function is evenly distributed over the chain.  
As an example, consider the chain with five sites which we can easily solve explicitly. In that
case
\begin{align}
\phi_0=\frac{1}{\sqrt{1+|t_1/t_2|^2(1+|t_3/t_4|^2) }}\left[ \begin{matrix}1 \cr0\cr  -t_1/t_2\cr 0 \cr (-t_3/t_4)(-t_1/t_2) \end{matrix}\right]
\end{align}
By appropriately engineering the ratios $t_1/t_2$ and $t_3/t_4$ we could place the zero mode at one edge: $$t_1<t_2,~t_3<t_4$$ the other edge:
$$t_1>t_2,~t_3>t_4$$ both edges: $$t_1<<t_2,(t_1/t_2)(t_3/t_4 )\sim 1$$ or the centre of the chain: $$t_1>>t_2,(t_1/t_2)(t_3/t_4 )\sim 1$$ Placing the support of the wave-function at both edges of the chain is interesting as it is located non-locally, reminiscent of Kitaev's Majorana Fermions on a finite chain \cite{Kitaev}.   In this case, if we put $$t_1=t_4~,~t_2=t_3$$ with $$t_1<<t_2$$ we find the non-zero energy levels are
\begin{align*}
\pm t_1~,~~\pm\sqrt{t_1^2+2t_2^2 }
\end{align*}
the gap between the zero mode and the first non-zero mode is $t_1$, the smaller of the two dimensionful scales in the problem.  
However, one could imagine that, in this small small system it might be possible to engineer the system so that $ t_1$ is larger than
the characteristic energies of processes which would affect the zero mode.

  In the following, we will concentrate on the case where the zero mode is located at an end of the chain.  The Su-Schrieffer-Heeger chain where $t_{2n}=t$ and $t_{2n+1}=t'$ with $t\neq t'$,  is a beautiful example of such a situation which has the added nice feature that the zero mode is isolated in a gap whose magnitude is $  |t'-t|$.  We can clearly see that if $t_{2k}/t_{2k+1}=t/t'<1$ for all $k=1,...,(N-1)/2$, the wave-function is localized near $n=1$ and it decays exponentially $\exp\left( -n \ln \frac{t'}{t} \right)$ as we move away from the edge.
  
  Before we continue, let us mention that the continuum analog of the odd-sited chain is the continuum problem governed by the Dirac Hamiltonian
  \begin{align*}
  E \left[ \begin{matrix} u(x) \cr v(x) \cr \end{matrix}\right] = \left[ \begin{matrix} 0 & \frac{d}{dx}+m \cr
  -\frac{d}{dx}+m & 0 \cr \end{matrix}\right]  \left[ \begin{matrix} u(x) \cr v(x) \cr \end{matrix}\right]
  \end{align*}
  where $0\leq x\leq L$.   The analog of the odd-sited spin chain is this Dirac equation with the two boundary conditions $v(0)=0$ and $v(L)=0$. This boundary condition is sufficient for the current density, $j(x)=-iu^*(x)v(x)
  +iv^*(x)u(x)$ to vanish at the boundaries of the system, which is in turn sufficient for the Hamiltonian to be self-adjoint. 
  The zero-mode solution is
  \begin{align*}
  \phi_0(x)= \left[ \begin{matrix}\sqrt{\frac{m}{e^{mL}-1}}\exp\left(mx\right)\cr 0 \cr \end{matrix}\right]
  \end{align*}
   The existence of this zero mode does not depend on the sign of $m$, since $u(x)$ has open boundary conditions, this is always a solution. 
  
  On the other hand, the even-sited chain is the analog of this Dirac equation with the boundary conditions $u(0)=0$ and $v(L)=0$.  
  In this case, there is no zero energy mode at all.  To find a solution, we could begin with the Ansatz 
  \begin{align*}
  u(x)= \frac{1}{c}\sin kx
  \end{align*}
  which satisfies the first boundary condition.  Then the Dirac equation tells us that 
   \begin{align*}
  v(x)=\frac{1}{c}\frac{1}{E}\left( m\sin kx-k\cos kx\right)
  \end{align*}
 and that
 \begin{align*}
 E=\pm\sqrt{k^2+m^2}
 \end{align*}
 The boundary conditions are satisfied if $k$ obeys the transcendental equation
  \begin{align*}
  \tan kL =\frac{k}{m}
  \end{align*}
  This equation always has an infinite number of solutions which are the allowed wave-vectors. 
  To find a bound state, we must look for a solution with imaginary $k$, that is, a solution of 
  \begin{align*}
  \tanh \kappa L =\frac{\kappa}{m}
  \end{align*}
This equation always has a solution when $m>0$ and it does not have a solution when $m<0$.  
When $m>0$ and $mL>>1$, $\kappa\sim m-2me^{-2mL}$ and $E\sim \pm 2me^{-mL}$.  When $m<0$, there are no states at all within
the energy gap. 
The continuum problem with $m<0$ is the analog of the even-sited SSH chain with strong bonds at each edge and the problem with $m>0$ is the analog of the even-sited SSH chain with weak bonds at each edge.  

\section{Zero modes and the quantum numbers of many-Fermion states}

We are interested in the influence of the zero energy modes on the many-body quantum states of electrons on
the odd-sited chain. Let us begin by studying the
lowest energy state of the system when it is charge neutral, that is, when it contains $N$ electrons, one for each of its $N$ sites.  
 We note that there are also $N$ energy levels, that is, there are $N$ eigenvalues of the single-particle Hamiltonian, $h$.   
 
 The occupation of the energy levels by electrons is governed by the Pauli principle.   The electrons have two spin-states and therefore two electrons
are allowed to occupy each energy level.  Let us assume that $N$ is odd. Then $N-1$ is an even integer.  The single particle energy levels consist of   $(N-1)/2$ positive energy states, $(N-1)/2$ negative energy states and one zero energy state.   A neutral state of the system will have $N$ electrons distributed amongst these states.    The lowest energy state with $N$ electrons has $N-1$ of those electrons forming  $(N-1)/2$ spin up-spin down pairs occupying the $(N-1)/2$ negative energy states.  Then the one additional electron must occupy the zero energy state.   This is the state of $N$ electrons which has the lowest possible energy.  It is charge neutral.  This is due to the fact that the electric charge of the $N$ electrons is canceled by the electric charge of the $N$ sites.  However, the state is two-fold degenerate, the zero energy state could have been occupied by a spin up or a spin down electron.  This leads to two distinct many-electron charge neutral ground states with the same energy.  They carry a doublet representation of the electron spin.   

This ground state has more degeneracy yet.  We could remove the electron which is in the zero energy mode.  This results in a state with charge $-e$, where $e$ is the electron charge. Since we de-populated a zero mode, the energy of the many-electron system is unchanged.  What is more, this state is  a singlet of the electron spin. The result is a state which has charge but no spin. Similarly, we could add an electron to the original neutral state so that there are then two electrons in the zero mode state.  These electrons would have paired spins, they would form a spin singlet and we would have a spinless state with charge $e$.  In this way we see that there are four degenerate ground states of the many-electron system, two of them neutral states which form a spin doublet and the other two are spin singlet states which have charges $e$ and $-e$. These are the unusual spin-charge quantum numbers which result from the presence of a single Fermion zero mode in the spectrum. 

\section{Entangled states}

\begin{figure}[h!]
\centerline{\includegraphics[scale=.8]{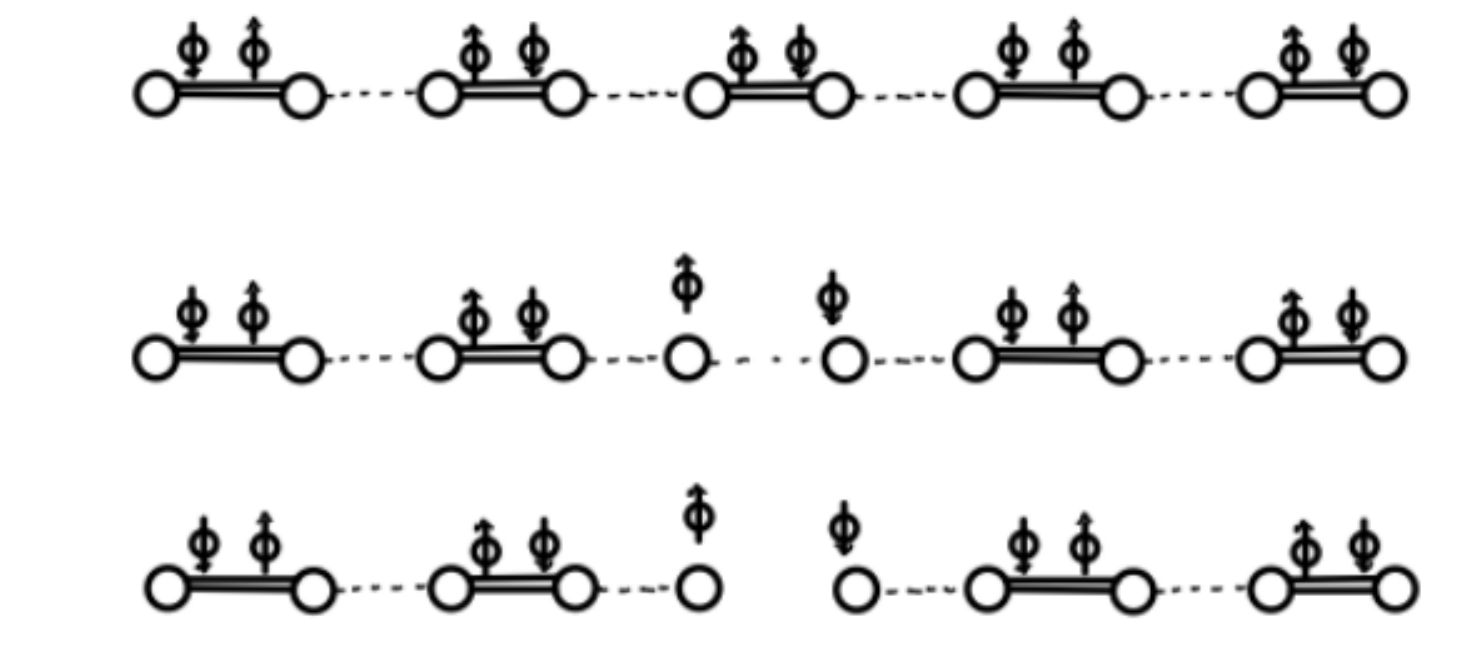}}
\caption{\small   Consider an even-sited Su-Schrieffer-Heeger chain where we adiabatically weaken a bond in such a way as to split it into two odd-sited Su-Schreiffer-Heeger chains.  If it was a strong bond, as is depicted in the figure, the zero modes of the two odd-sited chains which are produced reside on the edges of the odd-sited chains that are adjacent to the broken bond.  The resulting electronic state is a highly entangled state of the zero mode sectors on the two odd-sited chains. }
\label{figure3}
\end{figure}

We begin with an even-sited chain, where the strong bonds occur at the two edges of the chain, as is depicted in figure \ref{figure3}.  In this case, the electron spectrum on the chain is entirely gapped.  It has no energy states at all in the gap between $+|t-t'|$ and $-|t-t'|$. Then we split the chain into two odd-sited chains by adiabatically decreasing the one of the $t_n$'s to zero.  This must be a $t_n$ on a strong bond, since if it were on a weak bond the two remaining pieces of the chain would each have an even number of sites. The electronic state which is concentrated on that link contains a pair of electrons in a spin singlet.  The spin remains a singlet throughout the process. What is more, chain is charge neutral and the total charge remains at zero throughout.   To describe this system, we introduce two sets of creation and annihilation operators with the (non-vanishing) anti-commutation relations

\begin{align}
\left\{ a_{R,\sigma,0},a_{R,\sigma',0}^\dagger\right\}=\delta_{\sigma\sigma'}~,~
\left\{ a_{L,\sigma,0},a_{L,\sigma',0}^\dagger\right\}=\delta_{\sigma\sigma'}
\end{align}
one for the zero mode which arises on each of the odd-sited chains that are produced, which we can label as $R$ and $L$ for right and left.  
To understand what state is produced by the process, we could imagine a neutral state of the two odd-sited chains with the edges which support the zero modes brought into proximity.  They can interact simply by virtue of the fact that an electron can tunnel between them.  We can describe this by turning on an interaction Hamiltonian
\begin{align}
H_{\rm int}=\sum_\sigma \tau\left\{   a_{L,\sigma,0}^\dagger a_{R,\sigma,0}+
a_{R,\sigma,0}^\dagger a_{L,\sigma,0}\right\}
\end{align}
The charge neutral, spin singlet ground state of this Hamiltonian is
\begin{align*}
|{\rm gs}>~=~\frac{1}{2}\left( a_{L,\uparrow,0}^\dagger -  a_{R,\uparrow,0}^\dagger\right)
\left( a_{L,\downarrow,0}^\dagger -  a_{R,\downarrow,0}^\dagger\right)|0>
\end{align*}
where $|0>$ is the state of the two odd-sited chains where all of the zero modes are empty.
 This should be the state which is produced by turning off the link adiabatically.  This is a highly entangled state of the two chains.  To see this
 we could form the reduced density matrix for the $R$ chain by taking $|{\rm gs}><{\rm gs}|$ and tracing our the degrees of freedom of the $L$ chain.
 What we end up with is the unit matrix
 \begin{align*}
 \rho=\frac{1}{4}&\left\{ |0><0|+a_{R,\uparrow,0}^\dagger|0><0|a_{R,\uparrow ,0}
 +a_{R,\downarrow,0}^\dagger|0><0|a_{R,\downarrow,0} \right. \\
&\left.~~ +a_{R,\uparrow,0}^\dagger a_{R,\downarrow,0}^\dagger|0><0|a_{R,\downarrow,0}
 a_{R,\uparrow ,0}\right\}
 \end{align*}
 which has the maximum possible entanglement entropy for a four-level system, 
 $S_{\rm entanglement}=-{\rm Tr}\rho\ln\rho=\ln 4$. There exists an algorithm which uses the entanglement that is generated in this way to teleport the spin wave-function of an electron from the vicinity of one edge to the other edge \cite{Ghrear:2018yhb}.

\section{Spin dependent hopping}

 In the previous section, we reviewed the sense in which the Fermion zero mode leads to exotic quantum numbers of the many-Fermion system.  Because of the spin degeneracy, the states did not really have fractional charges, they  had  ``fractional charge per spin state'' which, with the two-fold spin degeneracy, resulted in the states having whole integer charges only.  It did  have the interesting upshot that the quantum numbers of the states were unusual, charge zero with spin 1/2 or charge $\pm e$ with spin $0$. 
 
 One might wonder whether we could find more interesting states if the Hamiltonian contained a spin-dependent term which removes the spin degeneracy.  An example would be a Zeeman interaction with a magnetic field which would shift the energies of one of the spin states to higher energies and the other spin state to lower energies.  This can be interesting and we will return to it, however, first, let us consider a more exotic possibility, one where the hopping amplitudes, $t_n$, of the tight binding model are spin-dependent.   We found in the last section that, generally, the Fermion zero mode is an edge mode if, generically, either $\frac{t_{\rm odd}}{t_{\rm even}}<1$ or $\frac{t_{\rm odd}}{t_{\rm even}}>1$, the zero mode has support at one or the other edge of the one-dimensional chain.  If the hopping amplitudes could be spin dependent in such a way that   $\frac{t_{\rm odd}}{t_{\rm even}}<1$ for spin up electrons and $\frac{t_{\rm odd}}{t_{\rm even}}>1$ for spin down electrons, the spin-up and spin-down zero modes would be localized at opposite edges of the one-dimensional chain.

To see the implications for charge density, let us examine the expectation value of the charge density in low energy overall neutral states. 
  The charge density of a given state is the expectation value of the operator, $\rho_n$, which measures the number of electrons at site $n$ multiplied by the electron charge $e$ 
  \begin{align}\label{density_00}
  \rho_n =   ~ea^\dagger_{\sigma, n} a_{\sigma ,n}~ ~-~e
  \end{align}
  where the last term are the charges of value $-e$ residing at each site of the chain (charges of the atoms when it is an atomic chain).
  The operator corresponding to the total charge is the sum of this charge density over the lattice sites, 
 \begin{align}
 Q=\sum_{n=1}^N \rho_n ~=~\sum_n e\left[ a^\dagger_{\sigma, n} a_{\sigma ,n}-1\right] 
  \end{align}
  The charge density obeys the continuity equation
  \begin{align}
  \frac{d}{dt}\rho_n ={\bf j}_n~-~{\bf j}_{n-1}
  \end{align}
 where the electric current ${\bf j}_n$ across the link between the site $n$ and the site $n+1$  is given by
  \begin{align}\label{current}
 {\bf  j}_n =it_n \left[  a^\dagger_{\sigma,n+1}a_{\sigma, n} -a^\dagger_{\sigma,n}a_{\sigma,n+1}\right]
  \end{align}
  
  If we consider the second quantization of the electron,
  \begin{align}\label{second_quantization}
  a_n = \sum_{E>0}\phi_{\sigma, E ,n}e^{-iEt} a_{\sigma ,E} +  \sum_{E<0}\phi_{\sigma, E ,n}e^{-iEt} b^\dagger_{\sigma ,E} +
  \phi_{\sigma, 0,n} a_{\sigma, 0}
  \end{align}
  where the non-vanishing anti-commutators are
  \begin{align}
  \left\{a_{\sigma, E}, a^\dagger_{\tau ,E'}\right\}=\delta_{EE'}\delta_{\sigma\tau}\\
  \left\{b_{\sigma, E}, b^\dagger_{\tau ,E'}\right\}=\delta_{EE'}\delta_{\sigma\tau}
  \end{align}
  The wave-functions for each spin state  are orthonormal
  \begin{align}
  \sum_n \phi_{\sigma, E, n}^*\phi_{\sigma ,E', n}=\delta_{EE'}~,~\forall E,E'
  \end{align}
  and they obey a completeness relation  
  \begin{align}\label{completeness}
  \sum_E \phi_{\sigma, E, n}\phi^*_{\sigma, E, n'} =\delta_{n n'}
  \end{align}
  In equation (\ref{completeness}), the summation is over positive, negative and zero energy states. 
  
  Let us consider the ground state  where all of the negative energy states, for both spin states, are filled and one of the zero modes  (here the spin up zero mode) is filled and the other zero mode is empty.  We denote this state by $|\uparrow>$.  It is 
  defined as the state of the second quantized system with the properties 
  \begin{align}\label{vacuum_1}
  &a_{\sigma ,E}|\uparrow>=0~,~b_{\sigma, E}|\uparrow>=0~,~\sigma=\uparrow,\downarrow\\
  &a^\dagger_{\uparrow ,0}|\uparrow>=0~,~a_{\downarrow, 0}|\uparrow>=0 \label{vacuum_2}\\
  &<\uparrow|\uparrow>=1
  \end{align}
  We denote the other overall charge neutral state as $|\downarrow>$ and it is defined by  
  \begin{align}\label{vacuum_2}
  &a_{\sigma ,E}|\downarrow>=0~,~b_{\sigma, E}|\downarrow>=0~,~\sigma=\uparrow,\downarrow\\
  &a^\dagger_{\downarrow ,0}|\downarrow>=0~,~a_{\uparrow, 0}|\downarrow>=0 \label{vacuum_2}\\
  &<\downarrow|\downarrow>=1
  \end{align}
  The expectation value of the charge density in the state $|\uparrow>$ is given by 
  \begin{align}\label{density_0}
 <\uparrow| \rho_n|\uparrow> & = e\sum_{E<0 }\left[ \phi^*_{\uparrow, E, n}\phi_{\uparrow, E ,n}+e\phi^*_{\downarrow ,E, n}\phi_{\downarrow ,E, n}\right] + 
  \phi^*_{\uparrow ,0, n}\phi_{\uparrow, 0, n}-e
  \end{align}
  Now, we use the fact that the negative and positive energy states are related by a simple transformation $\phi_{\sigma, -E}=\Gamma \phi_{\sigma,, E}$
  or $\phi_{\sigma, -E, n}=(-1)^n\phi_{\sigma, E, n}$ to
  write the sum over negative energy states in equation (\ref{density_0}) as half of the sum over all non-zero mode states, 
  \begin{align}
 <\uparrow| \rho_n|\uparrow>= \frac{e}{2}\sum_{E\neq 0 }\left[ \phi^*_{\uparrow, E ,n}\phi_{\uparrow ,E, n}+\phi^*_{\downarrow, E, n}\phi_{\downarrow, E ,n}\right] + 
  e\phi^*_{\uparrow, 0, n}\phi_{\uparrow, 0 ,n}-e
  \end{align}
  We can then use the completeness relation (\ref{completeness}) with $n=n'$ to get 
 \begin{align}
  <\uparrow| \rho_n|\uparrow>= \frac{e}{2}\left[2- \phi^*_{\uparrow, 0, n}\phi_{\uparrow, 0 ,n}- \phi^*_{\downarrow, 0 ,n}\phi_{\downarrow, 0 ,n}\right]+ 
 e \phi^*_{\uparrow, 0 ,n}\phi_{\uparrow ,0, n}-e\\ 
 \end{align}
 which, upon canceling some terms, yields the expression 
 \begin{equation}  \boxed{~~ <\uparrow| \rho_n|\uparrow>~=~
  \frac{e}{2}  \phi^*_{\uparrow, 0, n}\phi_{\uparrow, 0, n}-\frac{e}{2}
     \phi^*_{\downarrow, 0 ,n}\phi_{\downarrow, 0, n} ~~ }
      \label{charge_distribution}\end{equation}
 In addition, 
 \begin{align}
 <\downarrow| \rho_n|\downarrow> = - <\uparrow| \rho_n|\uparrow>
 \end{align}
 and 
  \begin{align}
 <\uparrow| \rho_n|\downarrow>  = 0 = <\downarrow| \rho_n|\uparrow>
 \end{align}
   These are  our central results.  
   
    The charge density is concentrated equally (and with opposite signs) on the densities of the zero mode wave-functions for each of the spin polarizations.  What is interesting here is that they can be in two different locations. In the spin-dependent SSH model, which we present a complete solution of in Appendix B, 
  the zero mode density $\phi^*_{\uparrow ,0 ,n}\phi_{\uparrow, 0 ,n}$ has support that is localized near one edge of the chain and $ \phi^*_{\downarrow, 0 ,n}\phi_{\downarrow ,0 ,n}$ has support that is localized near the other edge of the chain.  This fact does not depend on the length of the chains, so in principle the separation of the locations can be very large.  The wave-functions of the zero modes decay exponentially with distance from the edge, so for a long chain they would have vanishing overlap.  Moreover, all of the other electronic energy levels have an energy gap, $|t-t'|$, so these zero modes can by isolated from the rest of the energy spectrum.

   One might wonder in what sense this has to do with  fractional charge, as the states that we have discussed are overall charge neutral.  They are simultaneous eigenstates of the Hamiltonian and 
   of the total charge operator $Q=\sum_n \rho_n$.   The eigenvalue of $Q$ is zero.  The expectation value of the charge density is indeed concentrated at the edges of the chain.  However,  the charge density $\rho_n$, or the charge density in a subregion of the chain,  in spite of being Hermitian operators which could in principle be diagonalized, do not commute with the Hamiltonian, so their eigenstates are not stationary but are an admixture of states with different energies.  If we imagine measuring the charge residing, say, on the first site, $n=1$ of the chain, the measurement collapses the wave-function into the space of states corresponding to an outcome of the experiment, that is, eigenstates of $\rho_1$, which are $e,0,-e$, which are integer charges.  After such a measurement, one would expect that, with the help of a little dissipation, the system relaxes back to the same neutral ground state and the measurement can be repeated.  The average of the outcomes of many such measurements should yield the expectation value of the charge density on the first site of the chain, that is, the value of $<\uparrow|\rho_1|\uparrow>$ in 
   equation (\ref{charge_distribution}) with $n=1$.

   There are other diagnostics of a charge distribution which would be sensitive to its asymmetry in the states $|\uparrow>$ or $|\downarrow>$.  For example, in an external electric field of strength $E$, the chain would experience a torque of magnitude 
   \begin{align}
   {\mathcal T} = E\alpha\sin\theta \sum_n n \left<\rho_n\right>
   \end{align}
   where $\theta$ is the angle between the field and the chain. The torque exerted on the chain in the states $|\uparrow>$ and $|\downarrow>$ would be in opposite directions.  
   
   In the limit where the chain is long and we consider states such that the chain resembles a semi-infinite wire, if the limit is taken in such a way that the Hamiltonian of the semi-infinite system is a self-adjoint operator, the charge of the half of the system is a time-independent and it and the Hamiltonian have simultaneous eigenstates.   The limit of a long chain and the charge operator for one half of it can be defined by a limiting procedure where we
   define a smeared charge $Q_f=\sum_{n=1}^Nf(n) \rho_n$ with a test function $f(n)$ which has support near one edge of the chain and then goes to zero  in the central region.  We then first take the limit $N\to\infty$ where the chain is infinitely long and then we take the limit $f(n)\to 1$.  This will yield a
   charge operator whose eigenvalues are fractional. This procedure was carried out in the context of a slightly different model, with fractionally charged solitons in reference \cite{Jackiw:1983uf}.
   
   Let us consider the analog of the odd-sited spin chains that we have been discussing in continuum Dirac equations.  In this case, the spin up and the spin down wavefunctions, which we still label as $\left[\begin{matrix} u_\sigma(x)\cr v_{\sigma}(x)\cr \end{matrix}\right]$,  have the boundary conditions $v_\sigma(0)=0$ and $v_\sigma(L)=0$ and they satisfy different equations
   \begin{align}
   E\left[\begin{matrix} u_\uparrow(x)\cr v_{\uparrow}(x)\cr \end{matrix}\right]
   =\left[ \begin{matrix} 0 & \frac{d}{dx}+m \cr -\frac{d}{dx} + m & 0 \cr \end{matrix}\right]
   \left[\begin{matrix} u_\uparrow(x)\cr v_{\uparrow}(x)\cr \end{matrix}\right] \\
   E\left[\begin{matrix} u_\downarrow(x)\cr v_{\downarrow}(x)\cr \end{matrix}\right]
   =\left[ \begin{matrix} 0 & \frac{d}{dx}-m \cr -\frac{d}{dx} - m & 0 \cr \end{matrix}\right]
   \left[\begin{matrix} u_\downarrow(x)\cr v_{\downarrow}(x)\cr \end{matrix}\right]
   \end{align}
   The mass term appears with opposite signs for the spin up and spin down wavefunctions. 
   Now, the zero mode wavefunctions are
   \begin{align}
   \phi_{\uparrow 0}=\left[\begin{matrix} \sqrt{\frac{2m}{e^{2mL}-1}}\exp\left(mx\right)\cr 0\cr \end{matrix}\right]
   \\
   \phi_{\downarrow 0}=\left[\begin{matrix} \sqrt{\frac{2m}{e^{2mL}-1}}\exp\left(m(L-x)\right)\cr 0\cr \end{matrix}\right]
 \end{align}
 The spin up zero mode is concentrated near $x=0$ and the spin down zero mode is concentrated near $x=L$.

   \section{Epilogue: The $\{ |\uparrow>,|\downarrow>\}$ system as a qubit}
   
    The two neutral states $|\uparrow>$ and $|\downarrow>$ are the charge neutral states of the many-Fermion system.  A transition from these states to a state with excited particles and holes requires energy input to overcome the gap, $|t-t'|$ in the spectrum.  There are two further states
    that are degenerate in energy with them, the state with both zero modes empty, 
    $$
    a_{\downarrow 0}|\downarrow>~{\rm or}~a_{\uparrow}|\uparrow>
    $$
     and the state with both zero modes full, 
     $$
     a^\dagger_{\downarrow}|\uparrow>~{\rm or}~-a^\dagger_{\uparrow}|\downarrow>
     $$
  A transition to one of these from either of the states $|\uparrow>$ or $|\downarrow>$   requires the chain to absorb or emit an electron.  If such processes are suppressed, the doublet $\{|\uparrow>,|\downarrow>\}$ forms an isolated and interesting two-state system.  A dynamical mixing of the  states $|\uparrow>$ and $|\downarrow>$ by perturbations to the SSH Hamiltonian would be suppressed is the chain is sufficiently long that the wave-functions of the zero modes which reside at either edge have negligible overlap.  The transition would then require an electron to be transferred from one end of the chain to the other, a process which would be inhibited by the fact that the bulk material in between is an insulator with an energy gap $2|t-t'|$. 
  The Zeeman interaction 
  \begin{align}
  H_{\rm int,z} = \sum_n \frac{1}{2} \mu B_za_n^\dagger \sigma_z a_n
  \end{align} would still work to implement a single particle $z$-gate as one it is turned on the phases of the spin up and spin down wave-functions evolve with opposite signs, 
  \begin{align}
  |\uparrow>\to e^{iB_zt/2}|\uparrow>~~{\rm and}~~|\downarrow>\to e^{-iB_zt/2}|\downarrow>
  \end{align}
  
  If the system were smaller, so that the zero mode wave-functions overlap, the Zeeman interactions  with magnetic fields perpendicular to the direction of the spin polarization
  \begin{align}
  H_{\rm int,x} = \sum_n \frac{1}{2} \mu B_x~a_n^\dagger \sigma_x a_n
  \end{align}
  or 
  \begin{align}
  H_{\rm int,y} = \sum_n \frac{1}{2} \mu B_y~a_n^\dagger \sigma_y a_n
  \end{align}
  could be used to implement the single qubit $X$- and $Y$-gates, respectively.  
 However, at the same time, a shorter chain is more vulnerable to accidental environmentally induced transitions between $|\uparrow>$ and $|\downarrow$.

\appendix\section{The SSH chain}
\label{SSH_appendix}

In this appendix we will present the explicit solution of the SSH model for an odd-sited chain. 
The SSH model is a particular one-dimensional tight-binding model where the hopping amplitudes  are given by
\begin{align}\label{SSH}
 ~t_{2n}=t,~ t_{2n+1}=t'  
 \end{align}
 so that the tight-binding Hamiltonian is
 \begin{align}
 H=\sum_{\sigma=\uparrow,\downarrow}\sum_{n~{\rm even}}^N\left\{ t' a^\dagger_{\sigma n+1}a_{\sigma n}+t' a^\dagger_{\sigma n}a_{\sigma n+1}
 +t a^\dagger_{\sigma n+1}a_{\sigma n}+t a^\dagger_{\sigma n}a_{\sigma n+1}\right\}
 \end{align}
   The equations for the wave-functions  are
 \begin{align} \label{ssh_schroedinger_equation}
&n~{\rm even}:~E\phi _{E, n}= 
t \phi_{E, n-1}+t' \phi_{E, n+1} \\
&n~{\rm odd}:~E\phi _{E, n}= 
t' \phi_{E, n-1}+t\phi_{E, n+1} 
\end{align}
with the boundary conditions $\phi_{E,0}=0$ and $\phi_{E,N+1}=0$.  If N is odd, both boundary conditions apply to the even sites
and the wave-functions on even sites must have the form $\sin \frac{\pi kn}{N+1} $ where $k=1,2,...,(N-1)/2$. This function has the property that it vanishes when $n=0$ and when $n=N+1$.   With this observation, 
 the explicit solution for the electron wave-functions and energies is
 \begin{align}
 &n~{\rm even}:~\phi_{ E,n}=\sqrt{\frac{2}{N+1}}\sin \frac{\pi kn}{N+1} \label{ssh_wf_1}\\
 &n~{\rm odd}:~\phi_{ E,n}=\sqrt{\frac{2}{N+1}}\frac{1}{E(k) }\left[ t\sin\frac{\pi k (n+1)}{N+1}+t'\sin\frac{\pi k(n-1)}{N+1}\right] \label{ssh_wf_2} \\
 &E(k)=\pm\sqrt{ t^2+{t'}^2+2tt'\cos\frac{2\pi k}{N+1} } \label{ssh_energies} ~~,~~
  k=1,2,...,(N-1)/2 \\
 &\phi_{ 0,2n}=0~,~~\phi_{ 0,2n+1} = \sqrt{ \frac{1-t'/t}{1-(t'/t)^{(N+1)/2} }}\left( -\frac{t'}{t} \right)^n \label{ssh_zm_1}
 \end{align}
 The wave-functions in (\ref{ssh_wf_1}) and (\ref{ssh_wf_2}) contain the integer parameter $k$.  There are $(N-1)/2$ values of $k$, each of which corresponds to a positive-negative pair of energy levels with the dispersion relation
 given in equation (\ref{ssh_energies}). This makes $N-1$ energy levels for each spin state.  Then, there is the additional zero mode whose 
 wave-function is given in equation (\ref{ssh_zm_1}). This adds up to the expected $N$ energy levels .

If on the other hand, $N$ is even, we could make the same ansatz for the even sites
as we do in equation (\ref{ssh_wf_1}).  This guarantees the boundary condition $\phi_{ E,0}=0$. 
However, the condition $\psi_{E,N+1}=0$ belongs to the odd sites and equation (\ref{ssh_wf_2}) with $n=N+1$
gives us the transcendental equation for $k$
\begin{align}\label{gap}
0=t\sin\left( \frac{\pi k(N+2)}{N+1}\right)+t'\sin\left(\frac{\pi kN}{N+1}\right)
\end{align}
The solutions of this equation are the allowed wave-vectors.

\section{The spin-dependent SSH model}
   \label{spin-dependent_SSH}

In this Appendix, we will present the solution of the spin-dependent SSH chain.  Again, this is a tight-binding model but where the hopping parameters differ for the different, up and down, spin polarizations,
 \begin{align}\label{spin_dependent_SSH}
 {\rm spin}~\uparrow ~~t_{2n}=t,~ t_{2n+1}=t'  \\
 {\rm spin}~\downarrow ~~t_{2n}=t',~ t_{2n+1}=t  
 \end{align}
 
 The tight-binding Hamiltonian is
 \begin{align}
 H=\sum_{n~{\rm even}}^N\left\{ t' a^\dagger_{\uparrow n+1}a_{\uparrow n}+t' a^\dagger_{\uparrow n}a_{\uparrow n+1}
 +t a^\dagger_{\downarrow n+1}a_{\downarrow n}+t a^\dagger_{\downarrow n}a_{\downarrow n+1}
 \right\}\\
 +\sum_{n~{\rm odd}}^N\left\{ t a^\dagger_{\uparrow n+1}a_{\uparrow n}+t a^\dagger_{\uparrow n}a_{\uparrow n+1}
 +t' a^\dagger_{\downarrow n+1}a_{\downarrow n}+t' a^\dagger_{\downarrow n}a_{\downarrow n+1}\right\}
 \end{align}
 
 This model still has a partial spin symmetry in that the number operators for the total number of up spins  and for the total number of  down spins commute with the Hamiltonian separately. The SSH model of the previous Appendix had $U(1)\times SU(2)$ global symmetry.  In the present case,, this symmetry has been reduced to $U(1)\times U(1)$.  
 
In this case,  the spin up and spin down electrons each have a zero mode and in fact the energy eigenvalues for the spin up and spin down components of the wave-functions are still identical, and are in fact identical to the SSH model of the previous Appendix. 

What has changed is the fact that the spin up and spin down electrons with a given energy no longer have identical wave-functions.   
In fact, in this highly symmetric situation, the wave-functions are related by 
the transformation \begin{align}
\phi_{\downarrow,E,n}=\phi_{\uparrow,E,N+1-n}
\end{align}
   Moreover, when $t\neq t'$, the spectrum of non-zero modes has a gap, it obeys $E\geq |t-t'|$ or $E\leq -|t-t'|$.  
   
   We shall assume that at charge neutrality the chain has a density of one electron per site.  Due to the energy gap for moble sites, the charge neutral chain is an insulator.
   
   Here, we are assuming that the structure of the chain, that is, the even-odd nature of the hopping amplitudes $t$ and $t'$ is rigid so that the polyacetylene solitons are not allowed.   What is exotic here is that the, when $t\neq t'$, the spin-up and the spin-down zero modes have spatially separated support.  Our arguments of the previous sections tell us that they live at opposite ends of the odd-sited chain.  It is easy to confirm this by finding the solutions, which we shall write below.
 
 The equations for the wave-functions of the spin-dependent SSH chain are
 \begin{align} \label{ssh_2_schroedinger_equation}
&n~{\rm even}:~E\phi _{\uparrow,E, n}= 
t \phi_{\uparrow,E, n-1}+t' \phi_{\uparrow,E, n+1} \\
&n~{\rm odd}:~E\phi _{\uparrow,E, n}= 
t' \phi_{\uparrow,E, n-1}+t\phi_{\uparrow,E, n+1} \\
&n~{\rm even}:~E\phi _{\downarrow,E, n}= 
t' \phi_{\downarrow,E, n-1}+t \phi_{\downarrow,E, n+1}\\
&n~{\rm odd}:~E\phi _{\downarrow,E, n}= 
t\phi_{\downarrow,E, n-1}+t' \phi_{\downarrow,E, n+1}
\end{align}
 The explicit solution for the electron wave-functions and energies is
 \begin{align}
 &n~{\rm even}:~\phi_{\uparrow,E,n}=\sqrt{\frac{2}{N+1}}\sin \frac{\pi kn}{N+1} \label{wf_1}\\
 &n~{\rm odd}:~\phi_{\uparrow, E,n}=\sqrt{\frac{2}{N+1}}\frac{1}{E(k) }\left[ t\sin\frac{\pi k (n+1)}{N+1}+t'\sin\frac{\pi k(n-1)}{N+1}\right] \\
  &n~{\rm even}:~\phi_{\downarrow,E,n}=\sqrt{\frac{2}{N+1}}\sin \frac{\pi kn}{N+1} \\
 &n~{\rm odd}:~\phi_{\downarrow, E,n}=\sqrt{\frac{2}{N+1}}\frac{1}{E(k) }\left[ t'\sin\frac{\pi k (n+1)}{N+1}+t\sin\frac{\pi k(n-1)}{N+1}\right] \label{wf_4}\\
 &E(k)=\pm\sqrt{ (t-t')^2+4tt'\cos^2\frac{\pi k}{N+1} } \label{energies} ~~,~~
  k=1,2,...,(N-1)/2 \\
 &\phi_{\uparrow,0,2n}=0~,~~\phi_{\uparrow,0,2n+1} = \sqrt{ \frac{1-t'/t}{1-(t'/t)^{(N+1)/2} }}\left( -\frac{t'}{t} \right)^n \label{zm_1}\\
&\phi_{\downarrow,0,2n}=0~,~~\phi_{\downarrow,0,2n+1}=\sqrt{  \frac{1-t/t'}{1-(t/t')^{(N+1)/2} }}\left( -\frac{t}{t'} \right)^n \label{zm_2}
 \end{align}
 The wave-functions in (\ref{wf_1})-(\ref{wf_4}) contain the integer parameter $k$.  There are $(N-1)/2$ values of $k$, each of which corresponds to a positive-negative pair of energy levels with the dispersion relation
 given in equation (\ref{energies}). This makes $N-1$ energy levels for each spin state.  Then, there is the additional zero mode for each spin, whose 
 wave-functions are given in equations (\ref{zm_1}) and (\ref{zm_2}). This adds up to the expected $N$ energy levels for each spin state.

\end{document}